\documentclass[preprint]{vgtc}               

\ifpdf
  \pdfoutput=1\relax                   
  \usepackage{graphicx}                
  \DeclareGraphicsExtensions{.pdf,.png,.jpg,.jpeg} 
\else
  \usepackage{graphicx}                
  \DeclareGraphicsExtensions{.eps}     
\fi%

\graphicspath{{figures/}{pictures/}{images/}{./}} 

\usepackage{microtype}                 
\PassOptionsToPackage{warn}{textcomp}  
\usepackage{textcomp}                  
\usepackage{mathptmx}                  
\usepackage{times}                     
\usepackage{cite}                      
\usepackage{tabu}                      
\usepackage{booktabs}                  
\usepackage{amsmath}
\usepackage{amssymb}
\usepackage{amsfonts}
\usepackage{wrapfig}

\newcommand*{\simg}[1]{%
    \raisebox{-.3\baselineskip}{%
        \includegraphics[
        height=\baselineskip,
        width=\baselineskip,
        keepaspectratio,
        ]{#1}%
    }%
}

\onlineid{1186}

\vgtccategory{Research}

\title{Concept Lens: Visually Analyzing the Consistency of Semantic Manipulation in GANs}

\author{Sangwon Jeong\thanks{corresponding e-mail: sangwon.jeong@vanderbilt.edu}\\ %
    \scriptsize Lawrence Livermore National Laboratory %
\and Mingwei Li\\ %
     \scriptsize Vanderbilt University %
\and Matthew Berger\\ %
     \scriptsize Vanderbilt University %
\and Shusen Liu\\ %
     \scriptsize Lawrence Livermore National Laboratory %
     }

\teaser{
\centering
 \includegraphics[width=.85\textwidth]{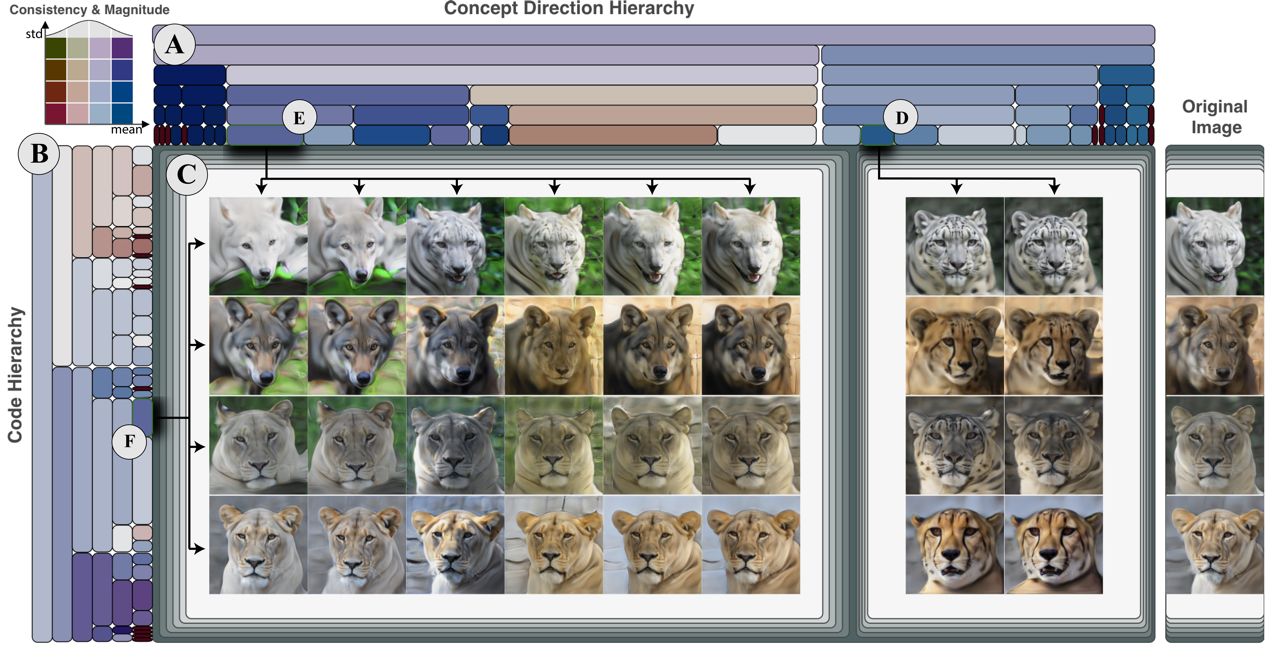}
 \caption{Concept Lens aims to support users in exploring and understanding the semantic manipulations offered by GAN latent spaces. The design of Concept Lens consists of (A) a hierarchical encoding of concepts, (B) a hierarchical encoding of codes, and (C) a grid-based view that shows images generated by combinations of codes (rows) and concepts (columns) selected by user. Nodes in the hierarchies color-encode the consistency and magnitude of edits made by concepts. In this example, conditioned on a group of codes at (F), the blue node (D) indicates consistent edits (e.g. the addition of spots), while the purple node (E) suggests inconsistent edits, shown as variability in species of animals.
 }
 \label{fig:teaser}
}

\abstract{
As applications of generative AI become mainstream, it is important to understand what generative models are capable of producing, and the extent to which one can predictably control their outputs. In this paper, we propose a visualization design, named \emph{Concept Lens}, for jointly navigating the data distribution of a generative model, and concept manipulations supported by the model. Our work is focused on modern vision-based generative adversarial networks (GAN), and their learned latent spaces, wherein concept discovery has gained significant interest as a means of image manipulation. Concept Lens is designed to support users in understanding the diversity of a provided set of concepts, the relationship between concepts, and the suitability of concepts to give semantic controls for image generation. Key to our approach is the hierarchical grouping of concepts, generated images, and the associated joint exploration.
We show how Concept Lens can reveal consistent semantic manipulations for editing images, while also serving as a diagnostic tool for studying the limitations and trade-offs of concept discovery methods.
} 

\keywords{Generative model, Explainable AI, Latent space, Clustering, Bivariate color}
\CCScatlist{
    \CCScat{}{Human Centered Computing}%
    {Visualization}%
}

\begin{document}


\firstsection{Introduction}
\maketitle

With recent research advancements \cite{brown2020language,ramesh2022hierarchical,saharia2022photorealistic}, generative AI has garnered tremendous interest.
For effective use of such models, however, two factors need to be considered: (1) the model's data distribution, e.g., what it is capable of producing, and (2) the level, and type, of control that a user has on the generation process. These are often competing interests, e.g., disentangled representations \cite{higgins2017beta,kim2018disentangling} offer semantic controls at varying levels of detail, yet are limited in modeling data, while diffusion-based models \cite{ramesh2022hierarchical,saharia2022photorealistic} generate highly realistic images but lack fine-grained controls, instead requiring tedious prompt engineering. 
For this reason, generative adversarial networks (GANs) remain popular, as they can generate highly realistic images \cite{karras2019style} while offering a more direct way to manipulate the output~\cite{wu2021stylespace,patashnik2021styleclip}. In GANs, generated images can be uniquely identified by a \emph{code} --- a point in the latent space of a GAN. 
Image manipulation can be achieved by moving the code along a direction in the latent space, and generating an image at the new location. 

Certain directions carry \emph{concepts}, where adding the concept direction to a code induces semantically meaningful change to the generated output. 
Numerous methods~\cite{harkonen2020ganspace,shen2021closed,yuksel2021latentclr} for discovering concepts have been developed, all with the goal of finding semantic edits, e.g. changing the facial expression of a person.

Despite the progress made in concept discovery, it remains a challenge to evaluate concept generality.
A direction can induce a consistent semantic change, but limited to a subset of the latent space \cite{jeong2022interactively,choi2021not}. 
For other portions of the latent space, either a different semantic change can occur, or simply no change is made. 
Although such heterogeneity has often been observed \cite{choi2021not}, a deeper understanding of this problem remains unexplored. 
Why, or why not, certain portions of latent space respond to concepts in a consistent manner? 
Answering this question can shed insight on how to better manipulate semantics in GAN latent spaces, while also uncovering potential dataset biases learned by generative models \cite{albahar2021pose,karakas2022fairstyle}.

We present Concept Lens, a visualization design for exploring and diagnosing concepts in GAN latent spaces. 
Concept Lens supports the analysis of arbitrary concepts, e.g. whether a concept is identified as a global direction \cite{harkonen2020ganspace,shen2021closed} or a vector field in the latent space \cite{yuksel2021latentclr}, and is domain-agnostic, e.g., we do not rely on a bank of external concepts for analysis. Further, unlike recent works that analyze a small set of latent dimensions~\cite{wang2023drava}, Concept Lens aims to relate a large (e.g. hundreds) set of concepts in order to provide a more complete picture of a GAN's latent space.

The generality of the problem under consideration, and the scale of the data, present challenges in analyzing concepts.
To this end, we employ hierarchical clustering of both concepts and codes as the basis for our exploration. 
Within the design, the nested hierarchies of concept and code enable a bi-hierarchy view of generated images (see Fig.~\ref{fig:teaser}). This design allows one to navigate the latent space at multiple levels of detail, individually for concepts and codes, and the relationship between them. Through linked brushing of the hierarchies, and a means to interactively steer clustering based on selected codes/directions, Concept Lens can reveal groups of codes that consistently respond to a given set of concepts, and vice versa -- what concepts lead to consistent semantic changes for a provided set of codes. Such information can be utilized to elicit advantages and limitations of existing concept discovery methods, improve semantic manipulation quality, and ultimately help the development of future methods.

We summarize the contributions of our work:
\begin{itemize}
    \item We present a visualization design that permits navigating the various types of semantics that can be manipulated in GANs.
    \item We introduce a bi-hierarchy visual encoding to help users relate latent space codes with concepts.
    \item We show how our design, termed Concept Lens, allows users to discover concepts that offer consistent semantic manipulations, as well as diagnose concept discovery methods and their potential trade-offs.
\end{itemize}

\section{Related Works}
\noindent\textbf{Concept Discovery Methods:} With recent developments \cite{karras2019style,karras2020analyzing} of high-quality GAN models, increasing research efforts have been dedicated to the unsupervised discovery and manipulation of concepts \cite{voynov2020unsupervised, harkonen2020ganspace, shen2021closed, yuksel2021latentclr, wu2021stylespace, patashnik2021styleclip, zhu2022region} in the generative output. 
Some of these methods \cite{harkonen2020ganspace, shen2021closed} try to identify directions of large variance in the internal representation of GAN models to uncover concepts, whereas others rely on external models \cite{patashnik2021styleclip} or local image space metrics \cite{wu2021stylespace, zhu2022region} for concept discovery.
These methods are further distinguished by their representation of concepts, e.g. a single global linear direction \cite{harkonen2020ganspace,shen2021closed}, a direction that depends on the latent code \cite{patashnik2021styleclip,yuksel2021latentclr}, or representing a concept as a curve in the latent space \cite{choi2021not,tzelepis2021warpedganspace}.
Concept discovery methods are, nevertheless, imperfect, and thus it is common to evaluate such methods on carefully-curated datasets \cite{eastwood2018framework,montero2022lost}, wherein the evaluation is presented as a set of summary statistics. However, quantitative metrics alone likely cannot provide full detail of the intricate behavior of semantic manipulation, whereas a visual analytics system can be valuable.

\noindent\textbf{Latent Space Visualization:} A few existing efforts have been proposed to study latent spaces using visualization.  Despite adopting different approaches, the Latent Cartography \cite{liu2019latent}, GANzilla \cite{evirgen2022ganzilla}, GANravel~\cite{evirgen2023ganravel}, and Drava \cite{wang2023drava} introduce interactive systems for exploring concepts in latent space by leveraging human annotation and feedback.
These methods are primarily focused on obtaining a single concept direction, rather than summarizing and exploring a large collection of concepts.
Recent work~\cite{jeong2022interactively} studies the disentanglement~\cite{eastwood2018framework} capabilities of concepts, yet visual analysis limited to a single concept.
A number of visualization works also focus on the comparison of latent spaces, such as embComp \cite{heimerl2020embcomp}, which proposes a visualization to compare multiple embedding at the same time, 
and Embedding Comparator \cite{boggust2022embedding}, which studies both the global and local structures of different latent spaces. 

\noindent{\textbf{Hierarchical Clustering in VA}: Hierarchical clustering and its visualizations \cite{wilkinson2009history, johnson1992treeviz, scheibel2020taxonomy, kruskal1983icicle} sits at the core of understanding tree-like structured data. Expandability of the hierarchical visualization allows for on-demand data exploration, contractability makes it a better overview of a very large dataset, such as biological or chemical data \cite{seo2002interactively}, compared to a flattened view such as scatter plots. For studying relationship between multiple datasets, using a single tree have limitations. Therefore, researchers developed a way to visualize a conjoined space to understand two distinctive data sets whether through merging \cite{yang2020interactive} or separately presenting them \cite{culy2010double, krause2023visually}.

\vspace{-3pt}
\section{Design Requirements}

We define the objectives of the proposed work, and outline design requirements.
We are interested in studying the space of images produced by a generator network of a GAN model~\cite{karras2020analyzing}. 
We assume a set of codes have been sampled in the GAN's latent space, where for each code we generate an image, and similarly a set of concept directions have been produced by a concept discovery method. These concept directions give users a way to control the output of the generator. We call a concept \emph{consistent} on given a set of codes, if when applying the concept direction one can observe the same semantic change in the output.

As has been studied in prior works~\cite{choi2021not,jeong2022interactively}, it is challenging for concept discovery methods to ensure consistency across all possible codes. In some cases this can be easily explained, e.g., for face images, it is unrealistic to expect a ``beard'' concept to readily apply to babies. In other instances, the root cause of inconsistency is more complicated to reason about, for instance owing to bias in the training data or limitation of the GAN. 
Relating a concept's inconsistency with the semantics of the latent space is essential for understanding concept manipulation, and GANs more broadly.

Another major challenge with state-of-the-art concept discovery methods \cite{voynov2020unsupervised, harkonen2020ganspace, shen2021closed, yuksel2021latentclr, wu2021stylespace, patashnik2021styleclip, zhu2022region} is the shear amount of concept directions they can produce. Despite the large quantity, these concepts are not labeled and often include many noisy and uninformative directions. How to understand, explore, and summarize them remains a challenge.
Therefore, \emph{the primary goal of the work is to address these challenges and support the exploration of consistency for a large number of direction-based manipulations in GANs.}
Specifically, we summarize the key design requirements \textbf{D1-D3} for supporting our objectives.

\textbf{D1 - Global summary of codes and concepts}: Since a large number of concepts will be generated when applying a concept discovery method, the system needs to support the summarization of all identified concepts as well as the manipulation results of applying them to a variety of codes, i.e., latent space samples.

\textbf{D2 - Estimate and communicate consistency}: Beside summarizing the exploration space, we also need to convey how similar a group of related concepts are; and how consistent a concept is when applying to different sets of codes.

\textbf{D3 - Focused analysis of local behavior}: Once we have a reliable way to express consistency, we can further our investigation through the local behavior of concept manipulation, e.g., 
contrasting different regions of the latent space by how they respond to a group of concepts.

\section{Concept Lens Design}

We detail the visualization design of Concept Lens, and how our design addresses the requirements previously discussed.

\noindent\textbf{Spatial Arrangement of Codes and Concepts}
\label{subsec:arrange}
The input to Concept Lens is a set of codes, and set of concept directions, where for each combination of code $\mathbf{w}_n \in \mathbb{R}^d$ and direction $\mathbf{d}_m \in \mathbb{R}^d$, we synthesize an image from a generator network denoted $G$, e.g. $G(\mathbf{w}_n + \alpha \mathbf{d}_m)$ is an image associated with direction $\mathbf{d}_m$ applied to code $\mathbf{w}_n$, with the scaling of the direction $\alpha \in \mathbb{R}$ representing the level of expression for the concept. For even modest-sized sets of codes (e.g. $N=500$) and concepts (e.g. $M = 500$), hundreds of thousands of images can result from this process.
To better organize the data, and allow for effective browsing at multiple levels of detail, we first perform hierarchical clustering of codes and concepts, individually, yielding two hierarchies. We use agglomerative clustering, with complete linkage, and permute the ordering of leaves via the method of Bar-Joseph et al. \cite{bar2001fast}. We visually encode each hierarchy as an icicle plot \cite{kruskal1983icicle}, with the concept-specific hierarchy arranged horizontally (Fig.~\ref{fig:teaser}(A)), and the code-specific hierarchy arranged vertically (Fig.~\ref{fig:teaser}(B)).

The center view (Fig.~\ref{fig:teaser}(C)) of Concept Lens is composed of a series of Cartesian products between nodes in the concept hierarchy and nodes in the code hierarchy, where each Cartesian product is limited to nodes at equal depth in respective hierarchies.
\begin{wrapfigure}{l}{0.20\textwidth}
\vspace{-.7cm}
  \begin{center}
    \includegraphics[width=0.23\textwidth]{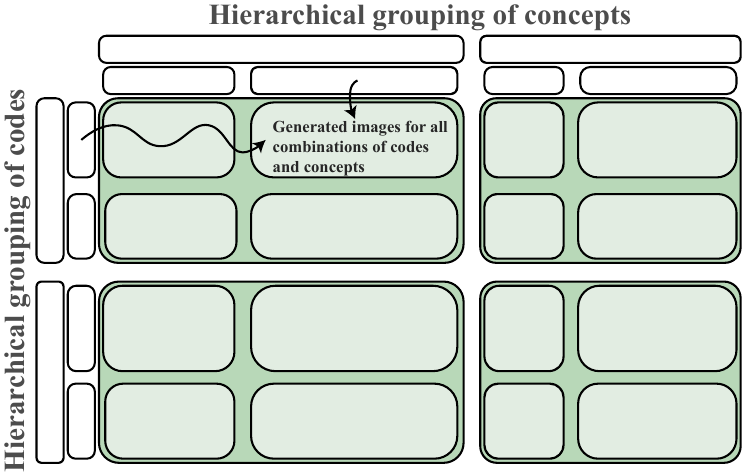}
  \end{center}
\vspace{-.7cm}
\end{wrapfigure}
For each resulting concept-code pair, a region of space is allocated, and a rectangle is positioned in this space whose size is proportional to node cardinality, e.g., its width proportional to the number of concepts in a node, and height proportional to the number of codes in a node -- this is depicted in the inset.
For a given pair of leaf nodes, we then display images in the allocated space for all combinations of its codes and concepts. 

Key to this design is the support for users to interactively select subsets of both hierarchies, via clicking or brushing, filtering images displayed in the center view reflecting selection in the hierarchies. Brushing gives a smooth navigation of codes and concepts (\textbf{D1}): pairs of leaf nodes that have a lowest common ancestor (LCA) deep in the hierarchy will be positioned close to each other~\cite{bar2001fast}. Moreover, to convey hierarchy depth, the design introduces gaps in the spacing between nodes, where the gap between two nodes is inversely proportional to their LCA depth. Such a design supports users in quickly obtaining a summary of the diversity of concepts (\textbf{D1}), as they can easily select concepts originating from distinct subtrees, and images of concepts from these subtrees will be displayed, with the hierarchy context preserved.

Building the aforementioned hierarchies requires a choice of distance measure to relate pairs of concepts, and pairs of codes.
In estimating the similarity between concepts, we need to take into consideration the variety of semantic shifts that can result from applying a concept to set of codes.
To this end, we propose a measure of \emph{aggregated concept distance}. Specifically, for a pair of concepts denoted $\mathbf{d}_j$ and $\mathbf{d}_k$, a distance measure is aggregated over all codes via:
\begin{equation}
    \vspace{-2mm}
    D(\mathbf{d}_j,\mathbf{d}_k) = \frac{1}{N}\sum_{n=1}^N \lVert \mathbf{f}_G(\mathbf{w}_n + \alpha \mathbf{d}_j) - \mathbf{f}_G(\mathbf{w}_n + \alpha \mathbf{d}_k) \rVert,
\label{eq:hierarchy}
\end{equation}
where $\alpha$ controls the strength of the edit and $\mathbf{f}_G$ computes a feature vector from the edited code; please see Appendix for feature choice and implementation details. 
The intuition behind this proposed measure is that two concepts should be reported as similar (low distance) if their respective edits to a single code are similar, summarized over \emph{all} ($N$ total) codes. The distance $D$ is provided as input for agglomerative clustering for concepts. For building the code hierarchy, we directly use the image feature $\mathbf{f}_G(\mathbf{w})$ of a code $\mathbf{w}$ to compute distances. 

\noindent\textbf{Measuring and Visualizing Consistency}
Although the bi-hierarchy view of Concept Lens aims to support fluid navigation of the GAN latent space, additional context is required for the user in making decisions on what to select and explore further. We address this challenge by measuring a notion of \emph{edit consistency} for groups of concepts, and groups of codes. Specific to a set of concepts denoted $S_D$ and a given code $\mathbf{w}$ we form the following collection of distances:
\begin{equation}
\mathcal{W}(S_D,\mathbf{w}) := \{ \lVert \mathbf{f}_G(\mathbf{w}) - \mathbf{f}_G(\mathbf{w} + \alpha \mathbf{d}) \rVert \, | \, \mathbf{d} \in S_D \}.
\label{eq:conceptconsistency}
\end{equation}
Each element in this set measures the feature-space distance between the original code ($\mathbf{w}$) and its corresponding edit ($\mathbf{w}+\alpha \mathbf{d}$). We measure code coherency similarly, conditioned on a set of codes $S_C$ and concept $\mathbf{d}$:
\begin{equation}
\mathcal{W}(S_C,\mathbf{d}) := \{ \lVert \mathbf{f}_G(\mathbf{w}) - \mathbf{f}_G(\mathbf{w} + \alpha \mathbf{d}) \rVert \, | \, \mathbf{w} \in S_C \},
\label{eq:codeconsistency}
\end{equation}

For concept consistency, the mean of the set $\mathcal{W}(S_D,\mathbf{w})$ gives a summary of edit magnitude, e.g., how much did the concepts in $S_D$ change the code $\mathbf{w}$? Likewise, the standard deviation of this set indicates variability amongst concepts in their edits, e.g., if two concepts share a semantic change, but one is entangled with an additional semantic, then this will manifest as high variability.

\setlength{\columnsep}{14pt}%
\begin{wrapfigure}{l}{0.22\textwidth}
\vspace{-.7cm}
  \begin{center}
    \includegraphics[width=0.24\textwidth]{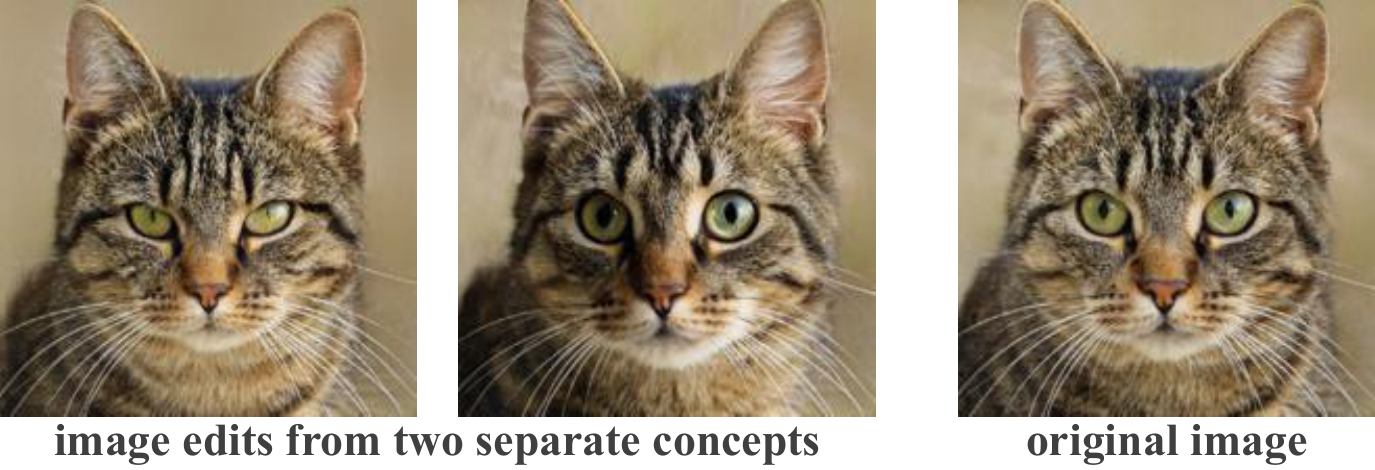}
  \end{center}
\vspace{-.7cm}
\end{wrapfigure}
This measure of \emph{edit consistency}, coupled with the \emph{aggregated concept distance} used for the clustering, serves as a proxy for \emph{semantic consistency} (\textbf{D2}), e.g., a cohesive group of concepts $S_D$ that yield edits of similar magnitude suggests semantic, predictable changes to a code.
In the inset we show an example of two concepts that, collectively, have moderate edit magnitude relative to the original code, but whose variability is high, e.g., shadows, and eyes closing.

\setlength{\columnsep}{14pt}%
\begin{wrapfigure}{l}{0.14\textwidth}
\vspace{-.7cm}
  \begin{center}
    \includegraphics[width=0.16\textwidth]{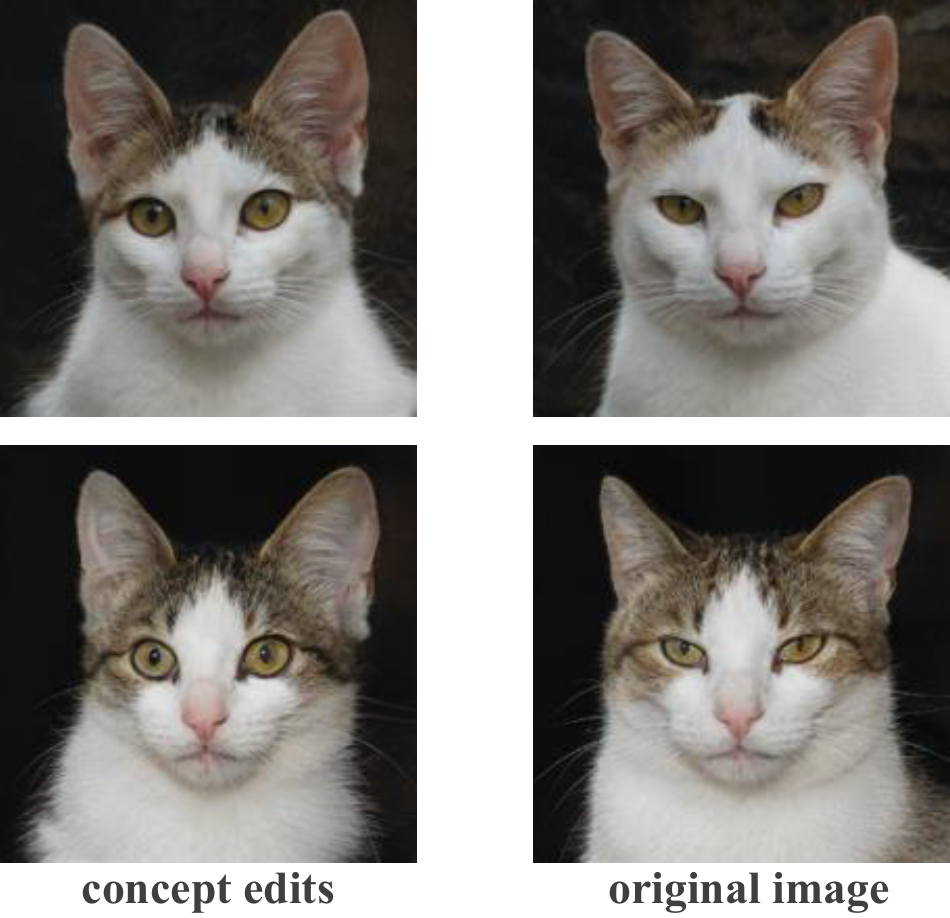}
  \end{center}
\vspace{-.7cm}
\end{wrapfigure}

For the case of code consistency, the summary statistics of the set $\mathcal{W}(S_C,\mathbf{d})$ indicate semantic consistency amongst a set of codes for a given concept. The inset shows an example wherein, for a single concept edit (left), and the original codes (right), the group of codes is measured as high edit magnitude, e.g., eyes opening and body pose change, while the variability between codes is small, e.g., the magnitude of change is consistent across codes.

We color map the consistency of edits over all nodes, across both hierarchies. Specific to the concept hierarchy, for a node with a set of concepts $S_D$, we compute the mean and standard deviation of $\mathcal{W}(S_D,\mathbf{w})$ for each code $\mathbf{w}$. When considering multiple codes, we average the two summary statistics over all codes, and visually encode the result with a bivariate color scale (see Fig.~\ref{fig:teaser}, top-left).
This scale is designed to support consistency analysis (\textbf{D2}) by emphasizing extremes, due to high saturation and low luminance, while mid-ranges between extremities are deemphasized via low saturation, attracting less attention.
For instance, red \simg{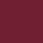} indicates concepts of small edit magnitude and low variability, blue \simg{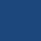} indicates edits that are large in magnitude yet low variability, while purple \simg{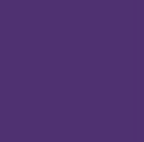} indicates both high edit magnitude and high variability. The domain for the scale is chosen based on the sample mean and standard deviation of \emph{edit consistency} computed on the entire dataset.





\begin{figure}[t]
  \centering
  \includegraphics[width=\linewidth]{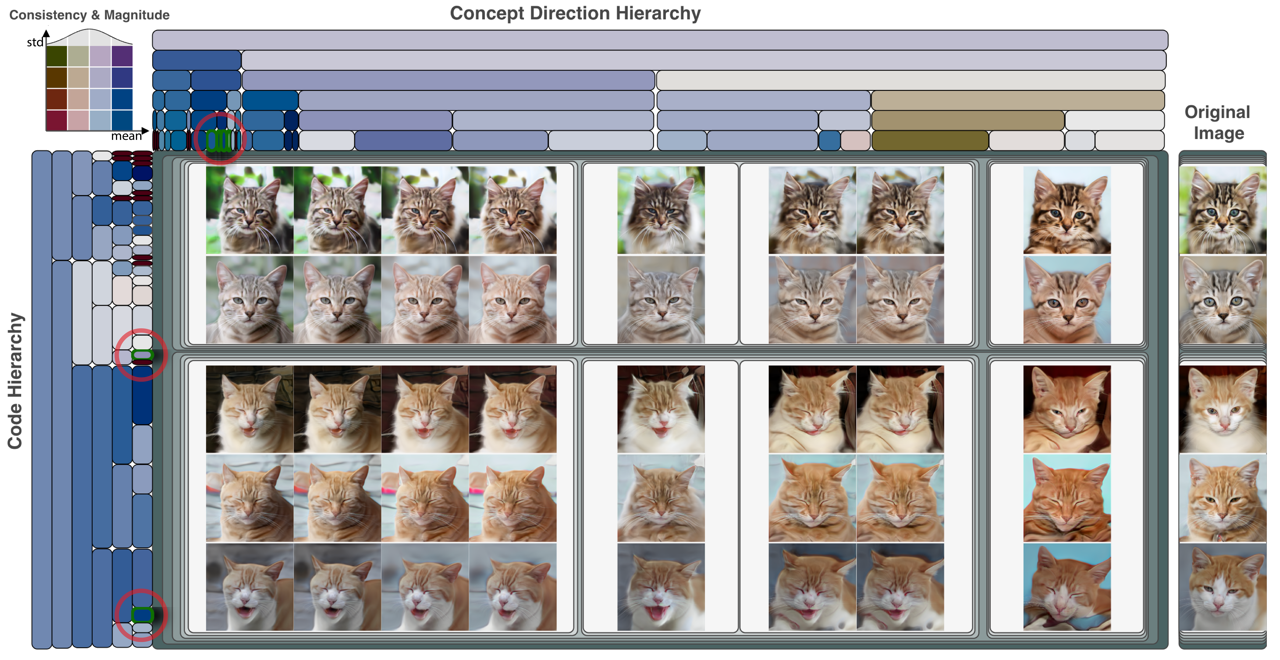}
  \caption{
  The selected concept applies ``closed eyes'' to most codes, while it doesn't to the top selection in the code hierarchy. The top group is kittens, and, in our dataset, we find that kittens usually do not close their eyes. For the cats in the bottom group, ``closed eyes'' concept is combined with ``mouth open''.}
  \vspace{-2mm}
  \label{fig:experiment-2}
\end{figure}

\noindent\textbf{Localized Analysis of Concepts}
Concept Lens further allows users to analyze consistency limited to subsets of codes, as well as subsets of concepts, to facilitate a more localized analysis of discovered concepts (\textbf{D3}).
Specifically, given a user's selection of nodes in the code hierarchy, we recompute edit consistency by limiting the computation in Eq.~\ref{eq:conceptconsistency} to the set of selected codes, and update the color of nodes in the concept hierarchy; a similar process is done in selecting directions, and updating the code hierarchy. This is designed to provide context for a user in what potential actions to take. For instance, a user might select a set of concepts, and judge the general semantics contained therein. Subsequently, the color-encoding of the code hierarchy will update, and from here a user might wish to see certain groups of codes that are consistent (blue) or have high variability (purple) (\textbf{D3}).

We further allow for the \emph{steering} of a given hierarchical clustering (e.g., concepts), based on a user's selection of the other hierarchy (e.g., codes) (\textbf{D3}). In the case of concepts, for a given set of codes selected by the user, we recompute concept similarity in Eq. \ref{eq:hierarchy}, only summing over the code selection. The updated concept hierarchy will thus be better organized for the provided code grouping, albeit at the expense of other codes. However, we view this action as a means of obtaining a more precise relationship between concepts, given a limited set of codes that a user is interested in. A similar process is done for steering the code hierarchy, given a set of concepts.
\section{Use Cases}

\noindent\textbf{Experiment Setup}
For our experiments, we use the StyleGAN2~\cite{karras2020analyzing} generator, SeFA \cite{shen2021closed} as our concept discovery method (targeting early layers in StyleGAN2), and consider several image domains: human faces, cat faces, faces of various animals~\cite{choi2020starganv2}, and portrait artwork (see video). Please see the Appendix for details on experiments.

\noindent\textbf{Exploration via Global Summary}
As a first step in exploration, it is necessary for the user to get a general sense of the types of images generated, the concepts found by discovery method, and the quality of these concepts. By browsing through the hierarchy, the user can identify the most representative codes and concepts by looking at examples from higher levels of the tree. Moreover, through the bivariate color encoding in the icicle plot (Fig.~\ref{fig:teaser}), at a glance the user can distinguish between consistent and inconsistent concepts as well as concepts enabling large edits.
The example shows the selection of two concept groups evaluated on one code group. The top-left (purple) and top-right (blue) concept selections differ in their respective levels of variability, while both giving large edits. Top-left concepts are more inconsistent, e.g. this concept group encodes the ``fox'' concept along with variations of colors, whereas the top-right group encodes only ``leopard''.

\noindent\textbf{Focused Investigation of Local Behavior}
Once users have obtained a high-level understanding of the data distribution and the variety of possible semantic manipulations, they can investigate the local behavior of concepts in regard to a specific code or group of codes.
In Fig. \ref{fig:experiment-2}, we see selection (top) captures a ``close eye'' concept for cats. This concept holds for most of the latent space, in line with the high consistency measure. Nevertheless, there are exceptions such as the nodes encoded with a lavender color. This signals that codes in this node change less when the concept is applied.
Here, we can observe that cats in the top group do not close their eyes, but those in the bottom group do, suggesting that the top group is less sensitive to the ``close eyes'' direction relative to other cats.
We also find the bottom group contains cats with their mouths open, suggestive of a potential dataset bias for orange-white colored Maine Coons.

\noindent\textbf{Reveal Unseen Pattern via On-demand Re-clustering}
Recall the concept hierarchy itself is built by aggregating pairwise distances over all codes. A consequence is that the hierarchy is not necessarily optimal for the exploration of a small group of codes, as covered in the previous use case. We can address this challenge through re-clustering.
On the human faces dataset, we observe a group of females with eyeglasses (c.f. ~Fig.\ref{fig:experiment-3-v2}). 
Upon selecting the group and browsing the original concept hierarchy computed over all codes, 
we find multiple small concept clusters scattered in the concept hierarchy that remove eyeglasses from faces. 
Limiting the distance aggregation to female faces with eyeglasses, and re-clustering, we find a group of large edits (blue colored node on the left of the concept hierarchy, Fig.\ref{fig:experiment-3-v2}) that remove the eyeglasses without modifying much of the face identities. The operation allows us to easily identify high-quality semantic manipulation for a focused group of codes.

\begin{figure}[t]
  \centering
  \includegraphics[width=\linewidth]{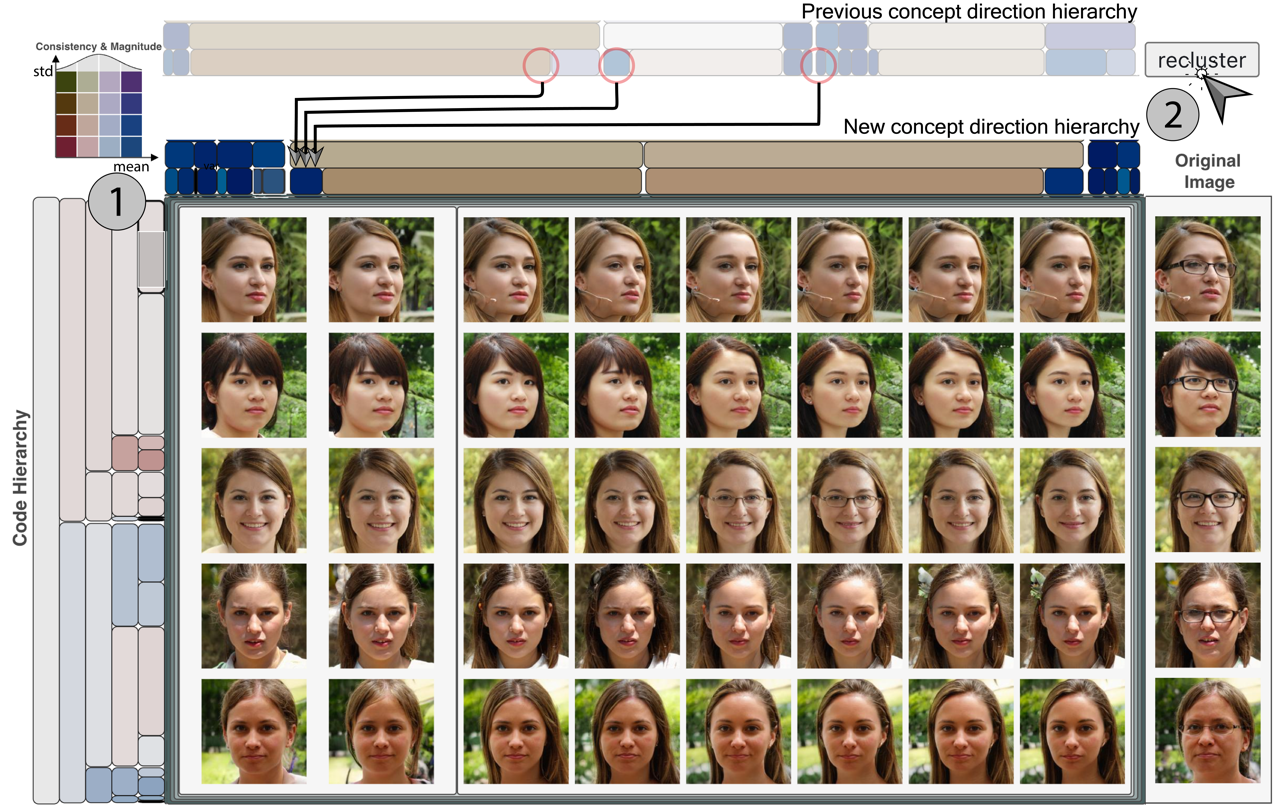}
  \caption{Re-clustering the concept hierarchy according to a set of closely related codes helps reveal the more consistent and high-quality concept edit. Re-clustering can be performed by brushing a subset of code in the code hierarchy (1) and clicking on the \textbf{recluster} button (2). Re-clustered directions are pointed by the arrows.}
  \label{fig:experiment-3-v2}
  \vspace{-2mm}
\end{figure}
\section{Discussion}
We introduce a visual analytics tool for investigating the consistency of semantics manipulation for applying concept directions generated by state-of-the-art concept discovery methods. 
With bi-hierarchical visual encoding in \emph{Concept Lens}, we showed that the proposed approach adequately addresses the scalability challenge thanks to the simplicity of required information (collection of original and edited images). Furthermore, the approach facilitates joint investigation between concept and code.
As part of future work we aim to apply Concept Lens to other types of GAN concept discovery methods, e.g. StyleCLIP~\cite{patashnik2021styleclip}, and other generative models, e.g. diffusion-based methods~\cite{saharia2022photorealistic}. Moreover, we plan to evaluate our design with users to assess its effectiveness in diagnosing concepts.

\acknowledgments{
This work was performed under the auspices of the U.S. Department of Energy by Lawrence Livermore National Laboratory under Contract DE-AC52-07NA27344. This work was supported by the Laboratory Directed Research and Development (LDRD) program under project tracking code 23-ERD-029 and reviewed and released under LLNL-CONF-848581.
}

\bibliographystyle{abbrv-doi}

\bibliography{template}

\appendix

\newpage
\section{Experimental details}

We provide more details on the data used in our experiments. We used StyleGAN2 \cite{karras2020analyzing} trained on varying domains, including FFHQ-dataset that contains high-quality human faces, cat faces, and faces of various animals.
For most experiments we randomly sampled 400 codes within the latent space of StyleGAN. For each code $\mathbf{w} \in \mathbb{R}^d$ we generate a corresponding image $G(\mathbf{w})$. In the interface these images are positioned to the right of the bi-hierarchy view, serving as a reference to compare against images generated by concepts.
Further, in most experiments we compute 400 concept directions via SEFA \cite{shen2021closed}. An image associated with code $\mathbf{w}$ is edited by a concept direction by computing $G(\mathbf{w} + \alpha \mathbf{d})$ for unit-norm concept direction $\mathbf{d} \in \mathbb{R}^d$. The parameter $\alpha \in \mathbb{R}$ represents the strength of the concept applied to the code; in practice, we set $\alpha = 5$.

To ensure that SEFA gives a diverse set of concept directions, we modify the method as follows.
We first collect weight matrices of affine transforms from StyleGAN that correspond to transforming codes from latent space to layer-specific style spaces. Namely, we gather weight matrices $\mathbf{A}_l \in \mathbb{R}^{o_l \times d}$ where $d$ is the dimensionality of input space and $o_l$ of the output space, for a given layer $l$. We limit our choice of layers to early in the generator network, to ensure large, structural changes to the output, rather than small color shifts that are usually associated with later layers. Moreover, to ensure diversity in concepts, we randomly drop rows of the affine matrices $\mathbf{A}_l$ to give $\mathbf{\tilde{A}}_l \in \mathbb{R}^{(o_l-r )\times d}$ where $r$ is the number of rows to drop, and stack the resulting set of matrices across all layers into one matrix, $\mathbf{\tilde{A}} = [\mathbf{\tilde{A}}_l]_{l=1}^{L}$. We then take the singular value decomposition of $\mathbf{\tilde{A}}$, and take the top $k$ (typically $k=20$) right singular vectors of largest singular values as the set of concept directions, though in practice we exclude the right singular vector with largest singular value, as it tends to give large transformations without meaningful semantics. This process is repeated with a different random sampling of rows, to given new matrices $\mathbf{\tilde{A}}$, and in turn new concept directions, until a target budget of concepts (e.g. 400) is obtained.

Lastly, image features are extracted from all images via ResNet fine-tuned for CLIP \cite{radford2021learning}. 
These image features are used for computing pairwise distances for hierarchical clustering, and concept and code consistency measures (Eq. \ref{eq:hierarchy}-\ref{eq:codeconsistency}). Image features $F \in \mathbb{R}^{k}$, where $k$ is the dimensionality of image feature space, need further processing for hierarchical clustering. Image features corresponding to every codes, applied by every directions $F \in \mathbb{R}^{C \times D \times k}$ needs centering to reduce the effect of starting position of codes in the clustering of walked codes. This is done by computing $F - \bar{F}$ where $\bar{F} = \sum_{i}^{C} \sum_{j}^{D} f_{i, j} / {C \times D}$. Cosine pairwise distance of these centered features are used for agglomerate clustering with complete linkage.



\end{document}